\documentclass{emulateapj}

\def\msini{$M_P\sin{i}$}

\def\msun{$M_{\odot}$}
\def\mjup{$M_{\rm Jup}$}

\def\fe{[Fe/H]}
\def\mkvk{$\{V-K_S$, $M_{K_S}\}$}

\begin{document}

\title{On the Metal-Richness of M Dwarfs with Planets}    

\author{John Asher Johnson\altaffilmark{1} \& 
Kevin Apps\altaffilmark{2}
}

\email{johnjohn@ifa.hawaii.edu}

\altaffiltext{1}{Institute for Astronomy, University
  of Hawaii, Honolulu, HI 96822; NSF Astronomy and Astrophysics
  Postdoctoral Fellow} 
\altaffiltext{2}{75B Cheyne Walk, Horley, Surrey, RH6 7LR, United Kingdom} 

\begin{abstract}
Knowledge of the metallicities of M dwarfs rests predominantly on the
photometric calibration of Bonfils and collaborators, which predicts
that M dwarfs in the solar neighborhood, 
including those with known planets, are systematically metal-poor
compared to their higher-mass counterparts. We test this prediction
using a volume-limited sample of low-mass stars,
together with a subset of M dwarfs with high-metallicity, F, G and K
wide binary companions. We find that the Bonfils et al. photometric
calibration systematically underestimates the metallicities of our
high-metallicity M dwarfs by an average of 0.32 dex. We derive a 
new photometric metallicity calibration and show that M dwarfs
with planets appear to be systematically metal-rich, a result
that is consistent with the metallicity distribution of FGK dwarfs
with planets.
\end{abstract}


\section{Motivation}
Mass and chemical composition play central roles in the birth,
evolution, and fate of stars. The growing ensemble of known exoplanets
discovered by Doppler surveys has revealed that the influences of
stellar mass and metallicity are not limited to the stars themselves,
but also 
extend to their planets. Abundance analyses of the stellar
samples encompassed by Doppler surveys show that the occurrence
rate of detectable planets correlates strongly with stellar
metallicity \citep{santos04, fischer05b} and stellar mass
\citep{lovis07, johnson07b}.   

In the standard core accretion model of planet formation the
relationships between stellar properties and the frequency of 
exoplanets are a reflection of the properties of the protoplanetary
disk. Stars with higher metallicities have disks containing a 
greater abundance of refractory material in the form of dust, allowing
rocky and icy cores to grow more rapidly \citep{ida04b}. Similarly,
the disks around 
massive stars also have enhanced surface densities, as well as
expansive radial regions in which protoplanetary cores grow most
efficiently \citep{laughlin04, ida05b, kennedy08}.    

The relationship between stellar mass and exoplanet occurrence was
first revealed by the paucity of Jovian planets around M dwarfs
\citep{endl03, butler06b, johnson07b}. However, unlike F, G and K
(FGK) dwarfs, accurate spectroscopic metallicities are difficult to
measure for M dwarfs because their spectra display 
complex and extensive molecular bands that leave no definable
continuum, and 
because of limited knowledge of the millions of transitions
responsible for the molecular absorption lines (Gustafsson
1989). The lack of spectroscopic metallicity measurements makes it
unclear whether stellar mass or metallicity represent the root 
cause of the dearth of Jovian planets around M dwarfs. 

Efforts have been made to calibrate the spectral features of low-mass
stars by inferring the metallicities of M dwarf secondaries from their
binary association with FGK primaries \citep[e.g.][]{mart08}, or by
the direct modeling of atomic lines from high-resolution spectra 
\citep{woolf06,bean06,maness07}. Unfortunately, these efforts have so
far been limited to a few specific stars and/or low metallicities.  

Another approach makes use of a broad-band photometric
calibration. The relationship between mass and absolute infrared
magnitude is very tight for the lower main sequence, with little
dispersion seen  beyond measurement errors \citep{delfosse00}. On the
other hand the relation between mass and absolute V-band magnitude
displays a large dispersion, and Delfosse et al. suggested that
metallicity variations as the cause of the observed scatter. This is
because increased stellar metal content leads to increased line
blanketing in 
the blue portion of the spectrum, causing metal-rich stars to move
redward in the H--R diagram such that they lie ``above'' the
main--sequence, e.g. more luminous than metal--poor stars of the same
color.

The scatter observed by Delfosse et al. is also clearly seen in the
dispersion about the main sequence in the \mkvk\ plane.
\citet[][hereafter B05]{bonfils05a} exploited the relationship between
stellar metallicity and distance away from a fiducial main sequence and
derived a broad-band, photometric ($V, K_S$) metallicity relation. B05
calibrated 
their polynomial relationship using the set of low-metallicity M
dwarfs in the  
\citet{woolf06} sample, as well as M dwarfs with FGK binary
companions. The metallicities of the latter sample were anchored to  
abundances derived spectroscopically for their corresponding FGK
binary companions, under the assumption that the binary components
are coeval and share the same chemical compositions.

The B05 calibration yields two remarkable results. First, M dwarfs
with planets appear to be metal-poor compared to
more massive FGK planet host stars. When applied to the 7 M dwarf
planet hosts  with accurate parallaxes, the B05 relationship yields a
mean \fe~$ = -0.11$, with only a single metal-rich host star,
Gl\,849. This result  
stands in stark  contrast to the sample of Sun-like stars with
planets, which have a mean [Fe/H]~$ =
+0.15.$\footnote{http://exoplanets.org} Second, for a  
volume-limited sample of 47 M dwarfs, 
B05 found a mean metallicity [Fe/H]~$ = -0.17$, which is 0.09 dex
lower than their volume-limited sample of FGK dwarfs. This implies
that the metallicities of 
low-mass stars in the Solar neighborhood are systematically lower than
the Sun-like stars, which leaves open the possibility that the decreased
frequency of Jupiters around M dwarfs is a reflection of their
lower metallicities rather than their lower masses.

B05 suggested that their observed metallicity offset between M dwarfs
and FGK stars is to be expected since M dwarfs have lifetimes
longer than the age of the Galactic thin disk \citep[e.g. $<
  9$~Gyr][]{sandage03, delpeloso05} and
should therefore trace the 
Galaxy's metal-poor past. However, K- and G-type stars
also have lifetimes longer than the age of the Galatic disk
\citep{hansen}, and yet show no 
metallicity offset compared to more massive, shorter-lived
stars \citep[cf. Figure 8 of][]{fischer05b}. To examine the
relationship between metallicity and spectral type further, we
compiled a representative sample of nearby, K2-G0 stars from the
Spectroscopic Properties of Cool Stars catalog
\citep[SPOCS;][]{valenti05}. To approximate a volume-limited sample we
used the 
absolute magnitude cuts described by \citet[][; $4.0 < M_V <
  6.5$]{reid07} and the distance limit ($d < 18$~pc) used by Fischer   
\& Valenti for their volume-limited subsamples of the SPOCS
catalog.

\begin{figure}[t!]
\epsscale{1.2}
\plotone{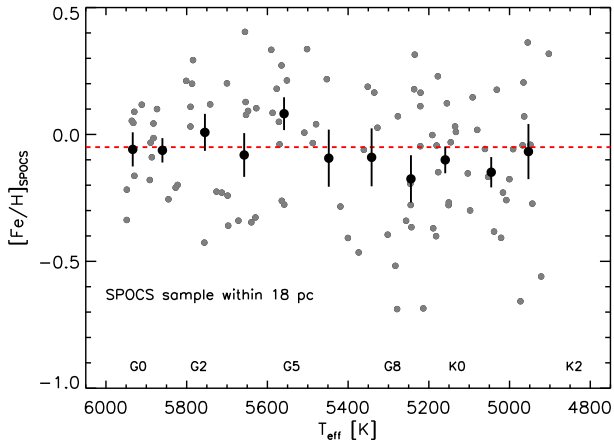}
\caption{The metallicity distribution as a function of effective
  temperature for all single stars in the SPOCS catalog with $d <
  18$~pc (gray circles). The large circles are the metallicities
  averaged within 100~K bins, and the error bars are the standard
  deviation of the mean [Fe/H] in each bin. The dashed line is the
  mean metallicity for the whole sample ([Fe/H]~$ = -0.05$). 
 \label{fig:feteff}} 
\end{figure}

Figure~\ref{fig:feteff} shows metallicity versus effective 
temperature (spectral type) for our volume-limited subsample of 109
stars. These stars, like M dwarfs, have
lifetimes longer than the age of the Galactic disk. Examination of the
metallicity distribution reveals no obvious decline in [Fe/H] with
decreasing $T_{\rm eff}$. We  calculated the Pearson correlation  
coefficient between [Fe/H] and $T_{\rm eff}$ and found 0.07, indicating 
little to no correlation between those properties among nearby
stars. Thus, there does not appear to be a physical reason for M
dwarfs to be systematically metal-poor compared to other stars in the
Solar neighborhood.

In this paper we investigate the validity of the photometric
calibration of B05 to ascertain whether the metallicity
offset for M dwarfs is real, or the result of a systematic error in
their photometric calibration. By addressing this issue, we aim to
disentangle the effects of mass and metallicity as the root cause of
the relative scarcity of Jovian planets around M stars.

\section{The $V-K$ versus $M_K$ Plane}

In this section we examine the B05 calibration using test stars
drawn from two volume--limited samples. The first sample is composed
of single M dwarfs within 10~pc and the second sample consists of
single, late K dwarfs within 20~pc. In what follows we use these two
stellar samples to measure the location of the mean main sequence for
low--mass stars in the solar neighborhood.

For our sample of K dwarfs we selected single stars with
Hipparcos-based parallaxes $\pi > 50$~mas (van Leeuwen 2007), parallax
uncertainties $< 5$\%, and
colors $2.8 < V-K_S < 3.5$. We obtained $V$-band magnitudes from
{\it Hipparcos}
and $K_S$-band magnitudes from the 2MASS catalog \citep{2mass}. Our
sample of M 
dwarfs comprises all known single stars with published parallaxes $>
100$~mas. Parallaxes and $V$-band magnitudes of our sample of M dwarfs
are primarily from the Hipparcos Catalog or the Yale Parallax Catalogue
\citep{yale}, and the $K_S$-band magnitudes are from  Hipparcos
and 2MASS, or in a few cases \citet{leggett92}, after applying
the transformation of \citet{carpenter01} to transform $K_{CIT}$ to $K_S$.  

\begin{figure*}
\epsscale{1}
\plotone{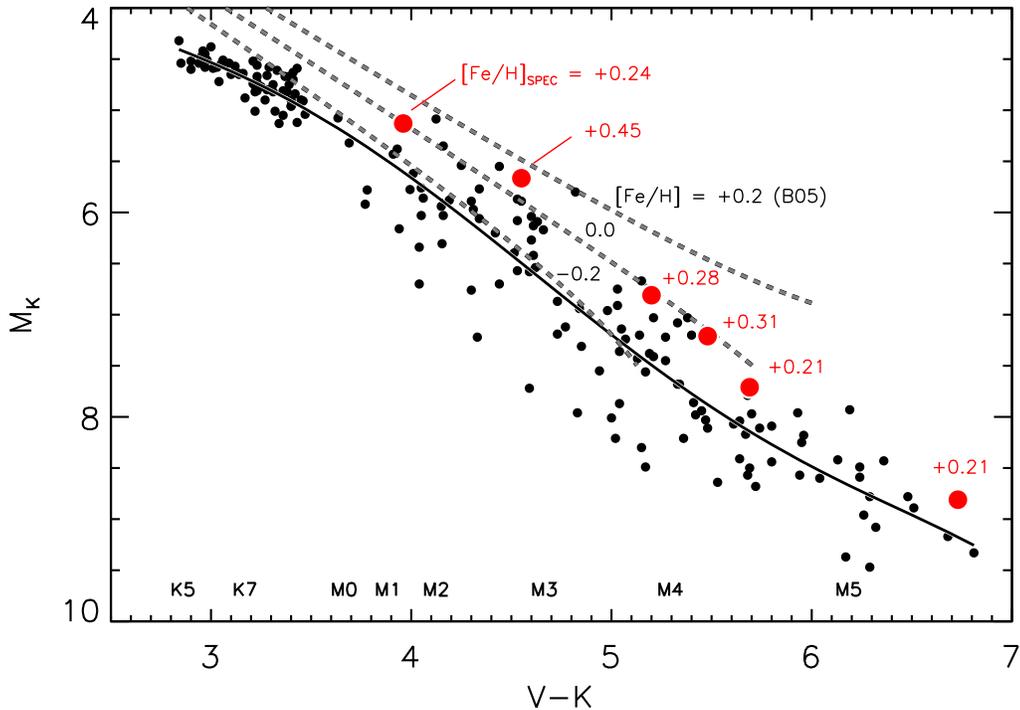}
\caption{Nearby low-mass stars in the \mkvk\ plane. The small black
  circles represent our volume-limited sample of single K dwarfs ($M_K
  \lesssim 5.5$, $d 
  < 20$~pc) and M dwarfs ($d < 20$~pc). The solid line is a
  fifth-order polynomial fit to the mean main sequence. The large
  filled circles are 
  the positions of a sample of high-metallicity M dwarfs, with
  spectroscopic \fe\ measured from their FGK binary
  companions. The $V-K_S$ colors of various spectral types are listed
  at the bottom of the figure. \label{fig:mk_vs_vk}} 
\end{figure*}

Figure~\ref{fig:mk_vs_vk} illustrates the \mkvk\ plane 
for our sample of late-type stars. Also plotted is our
fifth-order polynomial fit (solid line) that denotes the mean main
sequence for the Solar neighborhood, given by $\sum a_i (V - K_S)^i$
where $a = $\{-9.58933, 17.3952, -8.88365, 2.22598, -0.258854,
0.0113399\}. Also
shown are the 
isometallicity contours of the B05 calibration for 
[Fe/H]$ = \{-0.2,0.0, +0.2\}$, from bottom to top. Most of the stars
in our sample 
lie below the B05 solar-metallicity contour, illustrating the result
from B05 that most nearby M dwarfs are metal-poor.

We used the B05 calibration to derive 
metallicity estimates for all of stars within the color and
magnitude ranges over which the calibration is valid, namely $2.5 < V-K_S <
6.0$ and $4.0 < M_{K_S} < 7.5$. For the 66 stars that meet these criteria
we find a mean [Fe/H] of $-0.2$, with only 5 metal-rich stars stars 
($7.5 \pm 3.4$\% with \fe~$ > 0$). In contrast, $41 \pm 6$\% of the
K0-G0 dwarfs in our volume-limited subsample of the SPOCS catalog are
metal-rich (Figure \ref{fig:feteff}). Similarly, \citet{reid02} find
that $45 \pm 3$\%   
of FGK stars in the solar neighborhood have \fe~$ > 0$, and
\citet{haywood01} estimate that roughly half of the nearby K dwarfs
are metal-rich. Thus, applying B05 metallicity relationship to our
sample of nearby M dwarfs suggests a large abundance offset
between solar-type dwarfs and M stars.

\subsection{Testing the B05 calibration}

Also shown in  Figure~\ref{fig:mk_vs_vk} is a separate set of 
M dwarfs that have wide, common-proper-motion FGK binary
companions with precise spectroscopic metallicity measurements (red
circles). These 
six FGK$+$M binaries were selected from the SPOCS catalog and have
Hipparcos parallaxes measured to  
better than 5\% \citep{hipp}; $V$-band magnitudes from Hipparcos, and
$K_S$ magnitudes from the 
the 2MASS catalog and elsewhere in the literature; 
and most importantly, all are metal-rich with \fe~$ > +0.2$,
assuming the metallicities of these M dwarfs match those of their FGK
binary companions. We designate these stars as our
``high-metallicity'' sample and list their properties in
Table~\ref{tab:highmet}. 

The positions of our high-metallicity M dwarfs in the
\mkvk\ plane form a locus that lies along 
the upper edge of our sample of M dwarfs in Figure~\ref{fig:mk_vs_vk},
nearly parallel to the 
mean main sequence (solid line). Notably, our high-metallicity sample
lies along the solar-composition B05 contour rather than above
the [Fe/H]~$ = +0.2$ contour as would be expected. The B05 calibration
therefore systematically underestimates the metallicities of these
six metal--rich M dwarfs.

As a more quantitative assessment of the B05 calibration using the
high--metallicity M dwarfs, 
we first restrict our analysis to the region of the \mkvk\ plane over
which the calibration is valid, namely $4 < M_{K_S} < 
7.5$ and $2.5 \leq V-K_S \leq 6.0$ (B05). Four out of our six
high-metallicity stars meet these criteria. B05 predict a mean
[Fe/H]~$ = 0.0$ for these four stars, which is 0.32 dex lower than the
mean  metallicities of the FGK binary companions. 
This large discrepancy suggests that the B05
calibration contains an systematic error $-0.32$ dex for 
[Fe/H]~$ > +0.2$.

\begin{figure}[t!]
\epsscale{1}
\plotone{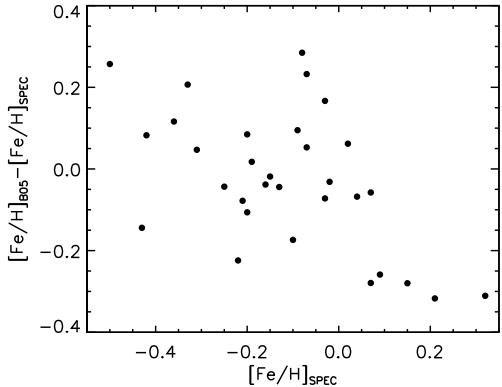}
\caption{The difference between the spectroscopic and photometric
  [Fe/H] for the sample of M dwarfs used to calibrate the B05
  metallicity relationship. The B05 relationship systematically
  underestimates the metallicities of all calibration stars with
  [Fe/H]~$ > 0$. \label{fig:fediff}}     
\end{figure}

We performed an additional test by examining whether the the B05
relationship could reproduce the 
spectroscopic metallicities of the stars in their calibration
samples. Figure~\ref{fig:fediff} shows the difference between the
spectroscopic and photometric \fe\ for the 19 calibration stars listed
in Table~3 of B05, and the 29 metal-poor stars from \citet{woolf06}.  
While the metallicities of the metal-poor stars are reproduced
reasonably well, the values for the metal-rich calibration stars  
are systematically underestimated by as much as 0.33 dex. The
inability of the B05 calibration to match the metallicities of
their calibration sample is additional evidence for a large
systematic error in the metallicity relationship, particularly
for \fe~$ > 0$.

We do not know the exact cause of this systematic error.
One possible source may be the limited metallicity
and color ranges spanned by the B05 calibration stars. Their first
calibration 
sample comprises 29 late-type dwarfs that have spectroscopic
metallicity measurements from \citet{woolf06}. The second set of stars
used for the B05 calibration consisted of 19 late K and M dwarfs with
FGK primaries. Both of these samples are predominantly metal-poor:
only 3 of 29 stars in the \citet{woolf06} sample are metal-rich, and only 5
out of 19 stars among the second set have have [Fe/H]~$ > 0$. Further,
only one star in their entire sample, 55\,Cnc\,B, has [Fe/H]~$ > +0.1$.

Another possible source of error is in the use of "visual magnitudes" 
\citep{gj91}, rather than
Johnson $V$-band magnitudes, for 7 out of the 19 M dwarfs in the B05
sample. The Gliese Catalog visual magnitudes were in many cases 
measured from photographic plates and therefore have non-negligible
and poorly characterized uncertainties, especially when compared to
the precision of modern $V$-band photometry.  
We searched the VizieR online catalog \citep{vizier} for alternative
photometric measurements for these 7 stars, and located $V$-band
photometry for 6 of them, mostly from the TYCHO/{\it Hipparcos} and
TASS Mark IV Survey catalogs \citep{perryman97, tass}. The difference
between the $V$-band and the  
visual magnitudes for these 6 B05 calibration stars has a median
offset of 0.3 and an rms scatter of 0.7. This $V$-band magnitude offset
is in the correct direction to account for part of the systematic
error in the B05 calibration at high metallicities, but the 0.7 mag
scatter precludes the application of a simple correction factor.

\begin{figure*}
\epsscale{1}
\plotone{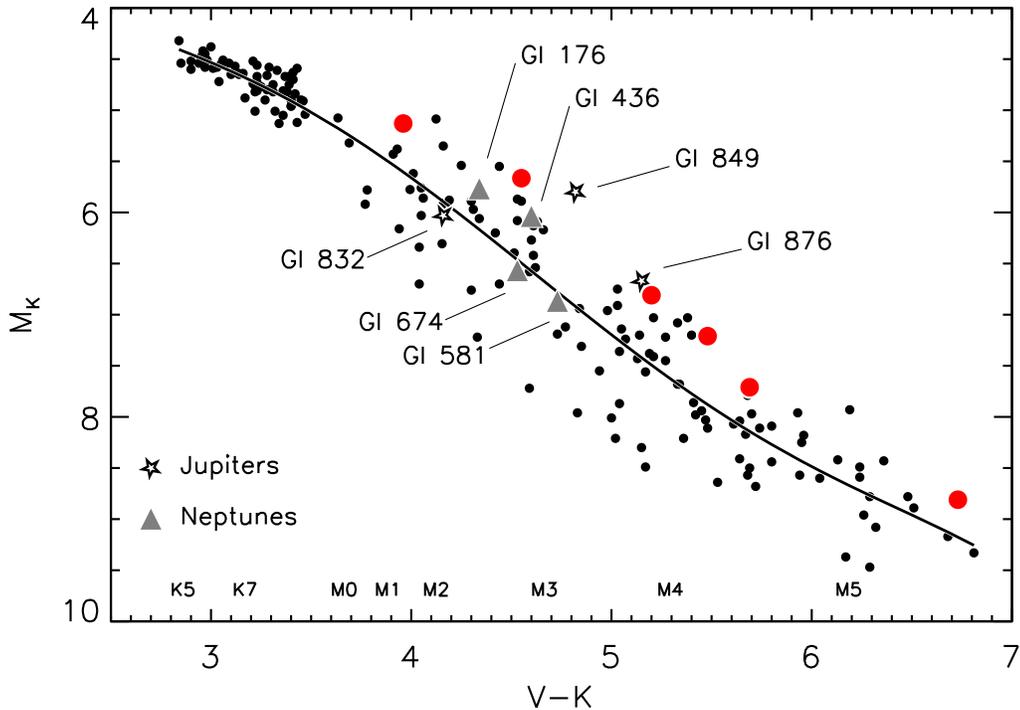}
\caption{Nearby low-mass stars in the \mkvk\ plane, along with M
  dwarfs known to harbor at least one planet. The small black
  circles a volume-limited sample of K and M dwarfs. The solid line is a
  fifth-order polynomial fit to the mean main sequence. The large
  filled circles are 
  the positions of a sample of high-metallicity M dwarfs, with
  spectroscopic \fe\ measured from their FGK binary
  companions. Five-point stars are M dwarfs with at least one Jovian
  planet, and the triangles are stars with at least one Neptune-mass
  planet. The $V-K_S$ colors of various spectral types are listed
  at the bottom of the figure. \label{fig:planets}}
\end{figure*}

\section{Examining the metallicities of M dwarfs with planets}

Based on the results in the previous section, we conclude that there
is a large systematic error in the B05 metallicity calibration,
particularly for [Fe/H]~$> 0$, and that
there does not appear to be an abundance offset between M dwarfs and
more massive stars. In light of this finding it is appropriate to
re-examine the metallicities of the known M dwarf planet host stars,
under the assumption that the metallicity distribution of stars in the
solar neighborhood is independent of stellar mass.

We begin by noting that the stars in our high-metallicity sample form a
well-defined locus in the \mkvk\ plane (Figure~\ref{fig:mk_vs_vk}),
and this locus lies parallel and to the right of the mean main
sequence. This suggests that the fundamental premise of the B05
calibration is valid: the metallicity of a given M dwarf should
correlate with its distance from the main sequence.

Figure~\ref{fig:planets} shows the 7 M dwarfs known to harbor
one or more planets. An eighth M dwarf planet host star, Gl\,317, is
omitted because it lacks a reliable parallax measurement
\citep{johnson07b}. Stars with Jovian planets are designated by
five-point stars, and those with Neptune-mass (\msini~$< 30 M_\oplus$)
planets are shown as triangles. All seven M dwarf planet host stars
lie near or above the mean main sequence in the \mkvk\ plane. Four of
the host stars---2/3 Jupiter-hosts and 2/4 Neptune-hosts---reside
above the main sequence and are therefore likely metal-rich. The
two Jupiter-hosts also lie above the
high-metallicity locus, indicating \fe~$ > +0.28$ based on the average
value of the high-metallicity sample. 

We now turn to a more quantitative approach to estimating the
metallicities of the M dwarfs with planets. As before, we assume the
fifth-order polynomial fit to the \mkvk\ distribution of low-mass stars 
represents an isometallicity contour with \fe$ = -0.05$, equal to the
mean \fe\ of the solar neighborhood. We then assume that \fe$ \propto
\Delta M_K$, where $\Delta M_K = MS - M_K$ is the distance away from the
mean main-sequence (MS) in the \mkvk\ plane. We found that a linear
relationship of the form 

\begin{equation}
{\rm \fe} = 0.56 \Delta M_K - 0.05
\label{fedv}
\end{equation}

\noindent produces the best results, yielding a dispersion 0.06
dex, based on the rms scatter of the difference between the predicted
and observed [Fe/H] for our set of high-metallicity
stars. Equation~\ref{fedv} is valid over the range $3.9 < V-K <
6.6$, which encompasses all known M dwarf planet hosts. 

We applied Equation~\ref{fedv} to the 7 M dwarfs with known planets
and precise parallaxes, and the results are given in the last column
of Table~\ref{tab:hostmet}. As in our qualitative analysis, we find
that 4 of the 7 M dwarf planet hosts are metal-rich, and of these
Gl\,849 and Gl\,876 have \fe$ > +0.3$. The three stars with sub-solar
metallicities are only slightly more metal-poor than the average
metallicity of the solar neighborhood---all 3 have \fe$ > -0.15$.
However, we note that we had to extrapolate beyond our data for the
region just below the main sequence. The
mean and median metallicities for the M dwarfs with planets are \fe$=
+0.16$ and $+0.19$, respectively, which is comparable to the mean
metallicity for the sample of known FGK dwarfs with planets
(\fe~$ =+0.15$).

\section{Summary and Discussion}

The results of Doppler-based planet searches have demonstrated that
giant planet occurrence correlates with stellar mass and
stellar metallicity. For FGK dwarfs, \fe\ is typically measured 
using LTE spectroscopic model fits to observed, high-resolution
spectra, yielding measurements with a precision of $\approx0.05$
dex \citep[e.g.][]{valenti05}. However, the complex spectra of
low--mass M dwarfs precludes the 
use of standard LTE spectral modeling, and knowledge of the
metallicity distribution of M dwarfs rests primarily on the
B05 photometric calibration. Based on their calibration, B05 reported
a systematic metallicity discontinuity for spectral types later than
$\sim$K7V, corresponding to stellar masses $M_* \lesssim
0.5$~\msun. This finding begs the question: is the deficit of Jovian
planets around M dwarfs due to their intrinsically low stellar masses
or metallicities?

We have addressed this question by studying several 
samples of low-mass stars in the \mkvk\ plane. Our volume-limited
sample of M dwarfs exhibits large scatter about the main
sequence, which is attributable to metallicity affects on the stars'
spectral energy distributions. Our second set of low-mass stars
consists of 6 M dwarfs with 
wide, FGK binary companions that have spectroscopic
metallicities [Fe/H]~$ > +0.25$ dex. 

We find that the B05 calibration underestimates the metallicities of
our high-metallicity stars by an average of 0.32~dex, and the B05
solar-abundance contour lies well above the mean main sequence of our
volume-limited set. Based on these results, we conclude that the systematic
metallicity offset for M dwarfs reported by B05 is a result of an
error in their calibration, rather than a reflection of the true
metallicity distribution of low-mass stars. It is therefore more likely
the decreased number of Jovian planets around M dwarfs is a
reflection of their lower stellar masses, rather than a 
metallicity effect.  

The cause of the systematic error in the B05 photometric metallicity
calibration is unclear. But the underlying concept, that metallicity
is directly proportional to height above the main sequence in the
\mkvk\ plane, is valid: our sample of 6 high-metallicity M
dwarfs form a locus that runs along the upper edge of the main
sequence in Figure~\ref{fig:mk_vs_vk}, offset from the mean main
sequence by $\Delta M_K \approx 0.5$. 
To estimate the metallicities of 7 M dwarfs known to harbor
planets, we derive a new photometric calibration based on a polynomial
fit to the mean main sequence of our volume-limited sample of low-mass
stars, and the locus defined by our set of high-metallicity
sample. Our calibration reveals that planets around M dwarfs are found 
preferentially around metal-rich stars, a result similar to the
metallicity bias observed for FGK planet host stars \citep{santos04,
  fischer05b}.

In addition to planet occurrence, stellar metallicity also correlates
with the physical properties of exoplanets. For example, the
relationship between minimum planet mass (\msini) and host-star
metallicity has  
recently received attention from several recent studies \citep{sousa08,
howard09}. Sousa et al. studied a large sample of stars with and
without planets and found a mean metallicity \fe~$ = -0.21$ for
stars that harbor at least one Neptune (\msini~$ \lesssim 0.1$~\mjup)
but do no Jupiters (\msini~$> 0.2$~\mjup). Their result 
implies that Neptune-mass planets form preferentially around
metal-poor stars, in contrast to stars with giant planets (mean
\fe~$+0.15$). They also 
noticed an  offset between the metallicities of stars with only
Neptunes compared  to stars with both Neptunian and Jovian planets,
which have a mean \fe~$ =  -0.03$. Sousa et al. summarized their
result as an observed increase in metallicity as a function of the
ratio of the number of Jupiters to Neptunes in a planetary system.

However, four out of the six stars in the Sousa et al. Neptune-only
sample were M dwarfs with metallicity estimates primarily from
B05. Using our metallicity estimates for the M dwarfs in their sample,
we find that their Neptune-only sample has a mean metallicity \fe~$ = 
-0.03$, which is equal to the average metallicity of stars with both
Neptunes and 
Jupiters. Thus, while the Neptune-hosts still appear to be metal-poor
compared to the Jupiter-hosts, there no longer appears to be a
metallicity offset between stars with Neptunes only and those with
both Neptunes and Jupiters. As additional low-mass planets are
discovered, the trends between stellar metallicity and the occurrence
of Neptunes and sub-Neptune-mass planets will come into better
focus \citep{mayor09, howard09}. 

Finally, it is important to caution the reader that our metallicity
calibration is meant only as a tool to gain a sense of the
metallicity distribution of M dwarfs with planets, and is currently
restricted to a limited color and magnitude range. For example, our
rough calibration was extrapolated to the region just 
below the mean main sequence. We feel this extrapolation was safe for
our study, but problems could arise if extrapolated further into the
metal-poor regime. We are currently obtaining spectroscopic
observations at Lick Observatory of a larger sample of FGK$+$M
binaries in order to derive a more thorough photometric calibration
extending over a broader region of the \mkvk\ plane.  

\acknowledgements

We gratefully acknowledge Geoff Marcy, Jon Swift, Mike Liu, Mike
Cushing, Andrew West and Jeff Valenti for their helpful conversations
and feedback. We also thank the anonymous referee for their helpful
feedback. JAJ is an NSF Astronomy and 
Astrophysics Postdoctoral Fellow with support from the NSF grant
AST-0702821. This publication makes use of data from the Two Micron
All Sky Survey (2MASS), which is a joint project of the University of
Massachusetts and the Infrared Processing and Analysis Center; 
the SIMBAD database operated at CDS, Strasbourge, France; and NASA's
Astrophysics Data System Bibliographic Services.

\bibliography{}

\clearpage

\begin{deluxetable}{llllllcc}
\tablecaption{High-metallicity M Dwarfs \label{tab:highmet}}
\tablewidth{0pt}
\tablehead{
  \colhead{Star Name}    &
  \colhead{Gliese} & 
  \colhead{Spectral} & 
  \colhead{$M_{K_S}$}    &
  \colhead{$V-K_S$}    &
  \colhead{$\Delta M_{K_S}$}    &
  \colhead{Spectroscopic}    &
  \colhead{B05 Calibration}    \\
  \colhead{}    &
  \colhead{Number} & 
  \colhead{Type} & 
  \colhead{}    &
  \colhead{}    &
  \colhead{}    &
  \colhead{[Fe/H]}    &
  \colhead{[Fe/H]}    
}
\startdata
 HD 46375 B  &         & M1   & 5.13 & 3.96 & 0.48 & $+0.24$ & $+0.00$ \\
 HD 38529 B  &         & M3   & 5.66 & 4.55 & 0.67 & $+0.45$ & $+0.03$ \\
 HD 18143 C  & 118.2   & M4   & 6.81 & 5.20 & 0.68 & $+0.28$ & $-0.02$ \\
 55 Cnc B    & 324 B   & M4   & 7.21 & 5.48 & 0.67 & $+0.31$ & $-0.01$ \\
 HD 190360 B & 777 B   & M4.5 & 7.71 & 5.69 & 0.43 & $+0.21$ &	$-0.03$ \\
 Proxima Cen & 551 C   & M5.5 & 8.81 & 6.73 & 0.36 & $+0.21$ & $+0.07$
\enddata
\end{deluxetable}

\begin{deluxetable}{clllllc}
\tablecaption{Properties of M Dwarf Planet Host Stars \label{tab:hostmet}}
\tablewidth{0pt}
\tablehead{
  \colhead{Gliese} & 
  \colhead{Spectral} & 
  \colhead{$M_{K_S}$}    &
  \colhead{$V-K_S$}    &
  \colhead{$\Delta M_{K_S}$}    &
  \colhead{[Fe/H]}   &
  \colhead{Planet}  \\
  \colhead{Number} & 
  \colhead{Type} & 
  \colhead{}    &
  \colhead{}    &
  \colhead{}    &
  \colhead{[Fe/H]}   &
  \colhead{Notes}
}
\startdata
876 & M4   & 6.67 & 5.15 & $+0.75$   & $+0.37$ & 2 Jupiters + Super-Earth \\
832 & M1.5 & 6.03 & 4.16 & $-0.13$  & $-0.12$ & Single Jupiter \\
849 & M3.5 & 5.80 & 4.82 & $+1.1$    & $+0.58$ & Jupiter + Trend \\
436 & M2.5 & 6.04 & 4.60 & $+0.54$   & $+0.25$ & Neptune + Linear Trend \\
581 & M3   & 6.87 & 4.73 & $-0.088$ & $-0.10$ & Neptune + 2 Super-Earths \\
674 & M2   & 6.57 & 4.53 & $-0.10$  & $-0.11$ & Single Neptune \\
176 & M2.5 & 5.77 & 4.34 & $+0.40$   & $+0.18$ & Single Neptune
\enddata
\end{deluxetable}

\end{document}